\documentclass[conference, a4paper]{IEEEtran}
\IEEEoverridecommandlockouts

\usepackage[turkish,english]{babel}
\usepackage[utf8]{inputenc}
\usepackage[T1]{fontenc}
\usepackage{cite}
\usepackage{color}

\ifCLASSINFOpdf
  \usepackage[pdftex]{graphicx}
\else
  \usepackage[dvips]{graphicx}
\fi
\graphicspath{{figures/}}
\usepackage[cmex10]{amsmath}
\usepackage{amssymb}
\usepackage{url}

\begin{document}
\shorthandoff{=}

\title{A Lightweight Method for Multiple Signal Direction Estimation with Adaptive Notch Filters}

\author{\IEEEauthorblockN{Burak Soner}
\IEEEauthorblockA{\textit{sobu Labs} \\
Çankaya, Ankara, Türkiye \\
burak@sobulabs.com}
}

\maketitle

\begin{abstract}
Capon beamforming, the conventional solution for multi-signal detection and direction-of-arrival (DoA) estimation, degrades in scenarios containing more transmitters than the number of receive antennas minus one. In this paper, a computationally lightweight method using adaptive notch filters (ANFs) is proposed for simultaneous DoA estimation of two or more narrowband signals with only two receive antennas. In the proposed method, ANF layers are applied to the incoming signal in a cascaded manner, and DoA estimation is performed via the Capon method on the isolated channels created for each signal. While the method offers very low computational load, in simulation results it exhibits performance similar to an ``oracle'' method that knows the signals in advance, for transmitters separable in the time-frequency-angle planes. As with ANFs themselves, the main limitation of this method is that multiple signals at closely spaced frequencies cannot be well separated. Alongside the simulation results, an experimental implementation on a low-cost software-defined radio platform (ADALM-Pluto) is also presented.
\end{abstract}

\IEEEpeerreviewmaketitle

\section{Introduction}

In transmitter localization and tracking applications, DoA estimation and triangulation methods using phased array antennas are widely employed \cite{vantrees2002}. In practical field conditions, multiple mutually independent transmitters broadcasting simultaneously are frequently encountered rather than a single transmitter. This problem has long been addressed in the phased array signal processing literature. Specifically, in the presence of a sufficient number of antennas, Capon beamforming \cite{capon1969} and similar methods can distinguish the directions of such multiple transmitters with high resolution \cite{schmidt1986}.

However, low-cost and embedded implementations of such systems with few antenna elements remain a current need, especially together with the recently widespread software-defined radio (SDR) platforms. In this paper, a computationally lightweight DoA estimation method based on adaptive notch filters (ANFs) is proposed, requiring only two receive antennas and targeting narrowband transmitters (constant or frequency-swept tones) that are sufficiently separable in the time-frequency-angle (space) planes. The main contributions of the paper are as follows:

\begin{itemize}
  \item The proposed method offers a lower computational load compared to Fourier transform based methods \cite{wang2014jsee}.
  \item For transmitters that are sufficiently separable in the time-frequency-angle planes, the method exhibits performance similar to an ``oracle'' method that knows the signals in advance.
  \item The method is experimentally demonstrated and analyzed through an implementation on the ADALM-Pluto SDR.
\end{itemize}

Alongside these contributions, it is shown that the inability to resolve multiple signals at closely spaced frequencies, the fundamental difficulty of ANFs, is also the main limitation of this method. Future work will be aimed at addressing this problem.

\section{System Model}

For an arrangement with \(M\) receive antennas, the superposition of \(N\) independent narrowband transmitters together with per-antenna additive noise is modeled as follows:
\begin{equation}
\mathbf{x}[n] = \sum_{i=1}^{N} \mathbf{a}_i(\theta_i)\, s_i[n] + \mathbf{w}[n],\quad \mathbf{x}[n]\in\mathbb{C}^{M}.
\end{equation}
Here, $\mathbf{x}[n]$ are the baseband signals at the receive antennas, the vector \(\mathbf{a}_i\) is the steering vector associated with direction \(\theta_i\), and \(s_i[n]\) is the baseband signal of the \(i\)-th transmitter. The noise \(\mathbf{w}[n]\) is assumed to be white (i.i.d.) across samples and elements.

For narrowband signals and steering vectors that are sufficiently separable across different angles of arrival, in an arrangement with \(M\) antennas the directions of at most \(M{-}1\) transmitters can be determined with classical Capon-type methods \cite{capon1969}. When the number of transmitters is \(N \ge M\), the performance of direct separation with Capon degrades (the estimate incorrectly converges to an intermediate solution between the directions of two transmitters) \cite{ma2010kr}. Classical subspace methods such as MUSIC and ESPRIT also rely on the same \(N<M\) assumption \cite{molaei2024review}. In this regime, it is possible to improve DoA estimation performance with a low computational load through a frequency-time plane decomposition such as the one proposed in this paper.

\section{Proposed Method}

In this section, the IIR ANF structure used for transmitter separation is first introduced, then the cascaded transmitter isolation stages built with ANFs are described, and finally the transmitter DoA estimation approach, applied on each isolated channel and realized via the Capon method, is explained. An important premise of the method is that it assumes the number of transmitters in the received signal is known in advance. This assumption can be satisfied with simple band-power detection additions to the existing cascaded ANF structure, but this feature is left outside the scope of the paper due to space constraints.

\subsection{Adaptive Notch Filter (ANF)}

ANF structures for tracking complex narrowband signals are widely known in the literature \cite{nehorai1985}. In practice, the single-pole IIR notch filter structure together with gradient-based optimization is preferred due to its low computational load \cite{borio2006ion}. With per-sample complexity of $\mathcal{O}(SM)$ for $S$ ANF stages and $M$ antennas, ANFs are more amenable to real-time operation on resource-constrained hardware than FFT-based methods, which run at least at $\mathcal{O}(M\log N)$ per sample \cite{trogliagamba2012}. Such an ANF tracks, sample by sample, the instantaneous frequencies of sufficiently separable narrowband components (constant tones and chirps), focusing on the dominant component at each successive stage.

In the proposed method, the ANF structure used for a single antenna is replicated identically for every antenna and every successive ANF stage. Given a complex sample \(x_n\), the filter notch frequency \(\omega_n\), and the pole-radius factor \(k_a\), the filter's intermediate state \(x^r_n\) and the filter output \(y_n\) take the form:
\begin{equation}
z_n = e^{j\omega_n},\quad
x^r_n = x_n + k_a z_n x^r_{n-1},\quad
y_n = x^r_n - z_n x^r_{n-1}.
\end{equation}
The notch frequency \(\omega_n\) is updated using the normalized least mean squares (NLMS) algorithm as follows \cite{haykin2013}:
\begin{equation}
\omega_{n+1} = \mathrm{clip}\!\left(
\omega_n - \frac{\mu}{\hat{P}_n}\,
\mathrm{Re}\!\left\{ y_n^* \, j z_n x^r_{n-1} \right\}
\right),
\end{equation}
where \(\hat{P}_n\) is the input power estimated via an exponential moving average (EMA), the NLMS gain \(\mu\) is a positive constant, and the clip operation is the bounding to the Nyquist limits.

\subsection{Signal Combining and Transmitter Isolation}

In the proposed method, for each antenna \(k\), an \(S\)-stage cascaded ANF is run on the baseband signal \(x_k[n]\). The output of one stage is the input of the next stage of the same antenna. At each stage, the per-antenna instantaneous frequency traces \(\hat{f}_k[n]\) (Hz) are combined across elements on a per-sample basis via arithmetic averaging:
\begin{equation}
\bar{f}[n] = \frac{1}{M}\sum_{k=1}^{M} \hat{f}_k[n].
\end{equation}
Once adaptation is complete, the combined traces \(\bar{f}^{(s)}[n]\), \(s=1,\ldots,S\), are applied to each antenna using the same fixed notch filter structure. Which transmitter physically corresponds to which stage is random, and the ordering need not be consistent. Which component is captured first depends on its dominance and on how close the current ANF happens to be to that band at the moment. For each target transmitter \(j\in\{1,\ldots,S\}\), notch filters with the combined traces of all stages with \(k\neq j\) are applied to the observed raw multi-channel instantaneous signal \(\mathbf{X}\), thereby producing the isolated signal channel corresponding to transmitter \(j\).

\subsection{Direction Estimation}

Once the isolated channels are obtained, for each channel an $M\times M$ sample covariance matrix \(\hat{\mathbf{R}}_j\) is computed. For a two-element ULA with steering vector \(\mathbf{a}(\theta)\), the Capon formulation is \cite{capon1969}:
\begin{equation}
P(\theta) = \frac{1}{\mathbf{a}(\theta)^H (\hat{\mathbf{R}}_j)^{-1} \mathbf{a}(\theta)}.
\end{equation}
The angle corresponding to the highest value (peak) in the Capon spectrum produced by this method is taken as the DoA estimate.

\section{Simulation and Experimental Results}

In this paper, to preserve realism in both open-air and indoor settings, the simulation accounts for path delay and multipath effects. To this end, a single-bounce model is used within a rectangular volume. In the environment summarized in Table~\ref{tab:simenv_en}, the two-element ULA center is placed at \((8.2, 2.8, 1.2)\,\mathrm{m}\), with element spacing \(\lambda/2\) (\(f_c=2\,\mathrm{GHz}\)), baseband sampling \(f_s=4\,\mathrm{MHz}\), instantaneous sample length \(N=512\), per-antenna noise \(\sigma=0.12~\text{pu}\) (on the scale used in the simulation), ANF parameters \(k_a=0.70\), \(\mu_{\mathrm{norm}}=0.1\), \(\alpha=0.01\) for EMA power estimation, and random-phase reflection coefficients in the range \(0.33\)--\(0.52\) on the six wall surfaces. All simulation results are obtained over \(400\) random Monte Carlo trials. Two main simulation scenarios are examined, along with a boundary analysis that measures the degradation under closely-spaced-frequency components:

\begin{table}[b]
  \vspace{10px}
  \centering
  \caption{Main parameters of the simulation environment}
  \label{tab:simenv_en}
  \begin{tabular}{|l|l|}
    \hline
    Quantity & Value \\
    \hline
    Room dimensions (m) & \(20\times 12\times 3\) \\
    Carrier frequency \(f_c\) & \(2\,\mathrm{GHz}\) \\
    Sampling \(f_s\), number of samples & \(4\,\mathrm{MHz}\),  \(512\) \\
    ULA element spacing & \(\lambda/2\) \\
    Noise (per antenna) & \(\sigma=0.12\) \\
    ANF (\(k_a\), \(\mu_{\mathrm{norm}}\), \(\alpha\)) & \(0.70\), \(0.1\), \(0.01\) \\
    Multipath & Single wall bounce, random phase \\
    \hline
  \end{tabular}
\end{table}

\begin{enumerate}
  \item \emph{Two transmitters:} results for two sources sufficiently separable in the time-frequency-angle planes, and comparison with the oracle method.
  \vspace{3px}
  \item \emph{Three transmitters:} a repetition of the same experiment with three transmitters, and observation of the negative effect of reduced separability.
  \vspace{2px}
  \item \emph{Boundary analysis:} an analysis of how the ANF traces mix as two transmitters approach each other and how this degrades DoA estimates, characterizing the limits of the proposed method.
\end{enumerate}

\begin{figure*}[t]
  \centering
  \includegraphics[width=0.94\textwidth]{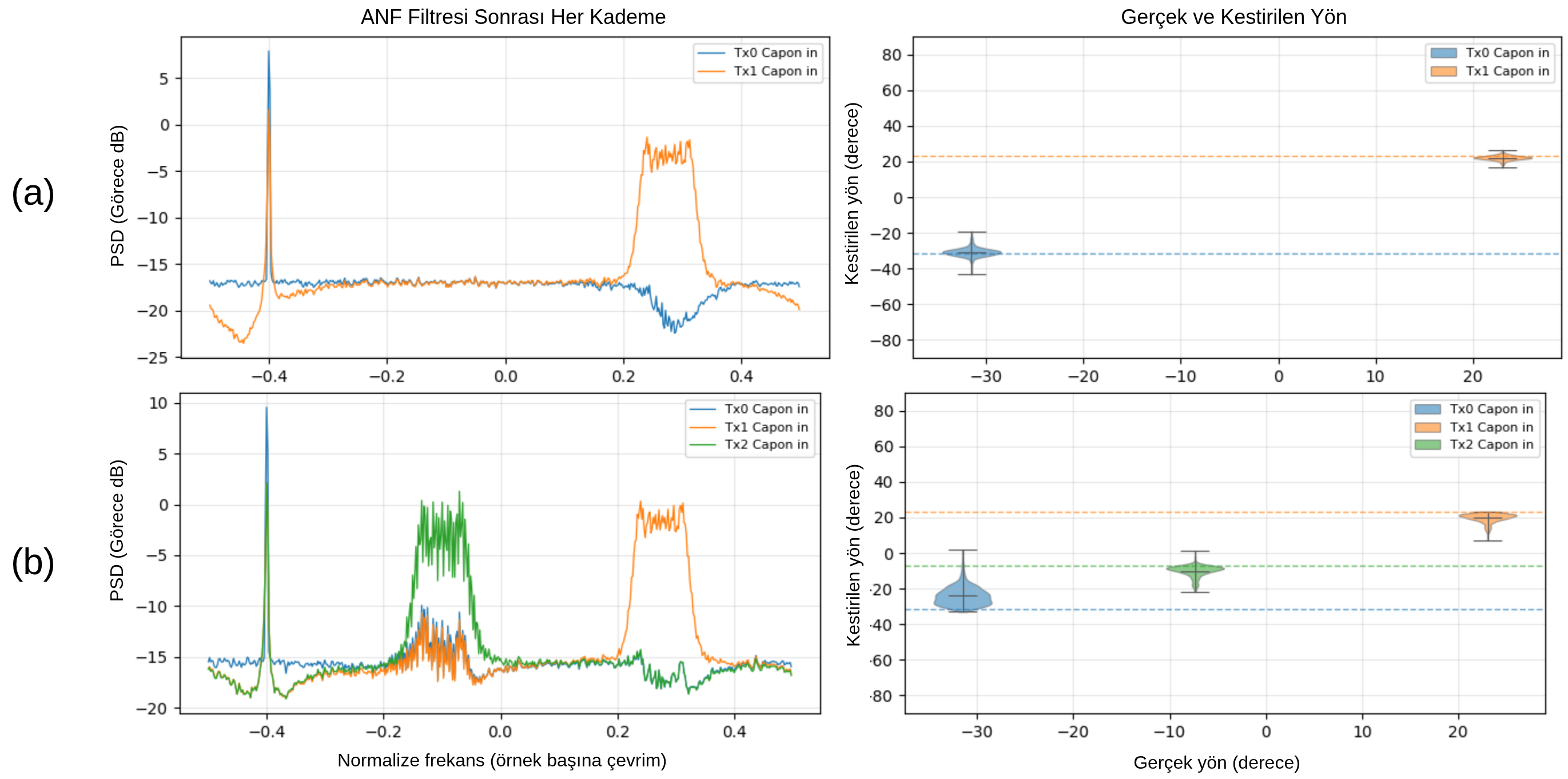}
  \caption{Simulation summary: (a) Two transmitters: post-filter channel PSD and Monte Carlo direction distributions (the Tx1 DoA estimate is more accurate than Tx0's), (b) Three separable transmitters: ANF+Capon outputs.}
  \label{fig:simall_en}
\end{figure*}

\begin{figure}[t]
  \centering
  \includegraphics[width=0.92\columnwidth]{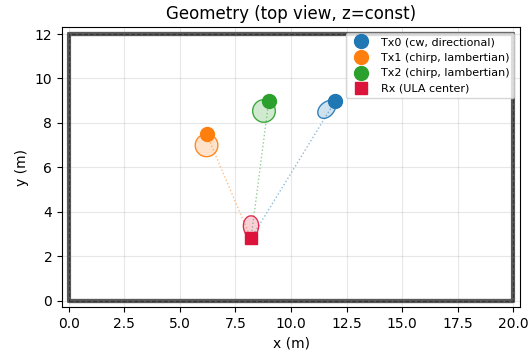}
  \caption{Simulation geometry (top view): three transmitters, the ULA center, and the room boundaries. Tx2 is disabled in the two-transmitter scenarios.}
  \label{fig:simsetup_en}
  \vspace{4px}
\end{figure}

\begin{figure}[t]
  \centering
  \includegraphics[width=0.92\columnwidth]{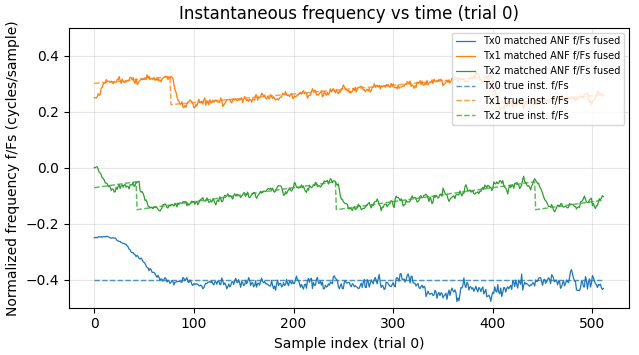}
  \caption{Instantaneous frequency trace of the three-stage ANF structure. After the initial offsets, the traces settle into the frequency regions corresponding to the physical transmitters.}
  \label{fig:sim3txanf_en}
  \vspace{7px}
\end{figure}

\begin{table}[t]
  \centering
  \caption{Boundary analysis for two transmitters with approaching frequencies}
  \label{tab:boundary_en}
  \begin{tabular}{|r|r|r|}
    \hline
    $\Delta f$ (kHz) & Tx0 RMSE & Tx1 RMSE \\
    \hline
    $-2500$ & 4.44 & 1.50 \\
    $-1200$ & 4.70 & 1.68 \\
    $-900$ & 5.56 & 2.26 \\
    $-750$ & 15.12 & 3.30 \\
    $-600$ & 24.66 & 6.44 \\
    $-200$ & 26.79 & 5.58 \\
    \multicolumn{1}{|l|}{Overlap} & 35.23 & 6.66 \\
    \hline
  \end{tabular}
\end{table}

Fig.~\ref{fig:simsetup_en} shows, from above, the three transmitters (Tx0--Tx2), the two-element ULA receiver, and the single-bounce multipath geometry on the simulation plane. All transmitters are kept at the same \(z\) height so that only azimuth DoA estimation is targeted with the two-antenna ULA. The antenna beams of the transmitters are depicted as elliptical (directional) and round (Lambertian) in the figure. In the two-transmitter experiments, Tx2 is turned off and only Tx0 and Tx1 are used.

\begin{figure*}[t]
  \centering
  \includegraphics[width=0.90\textwidth]{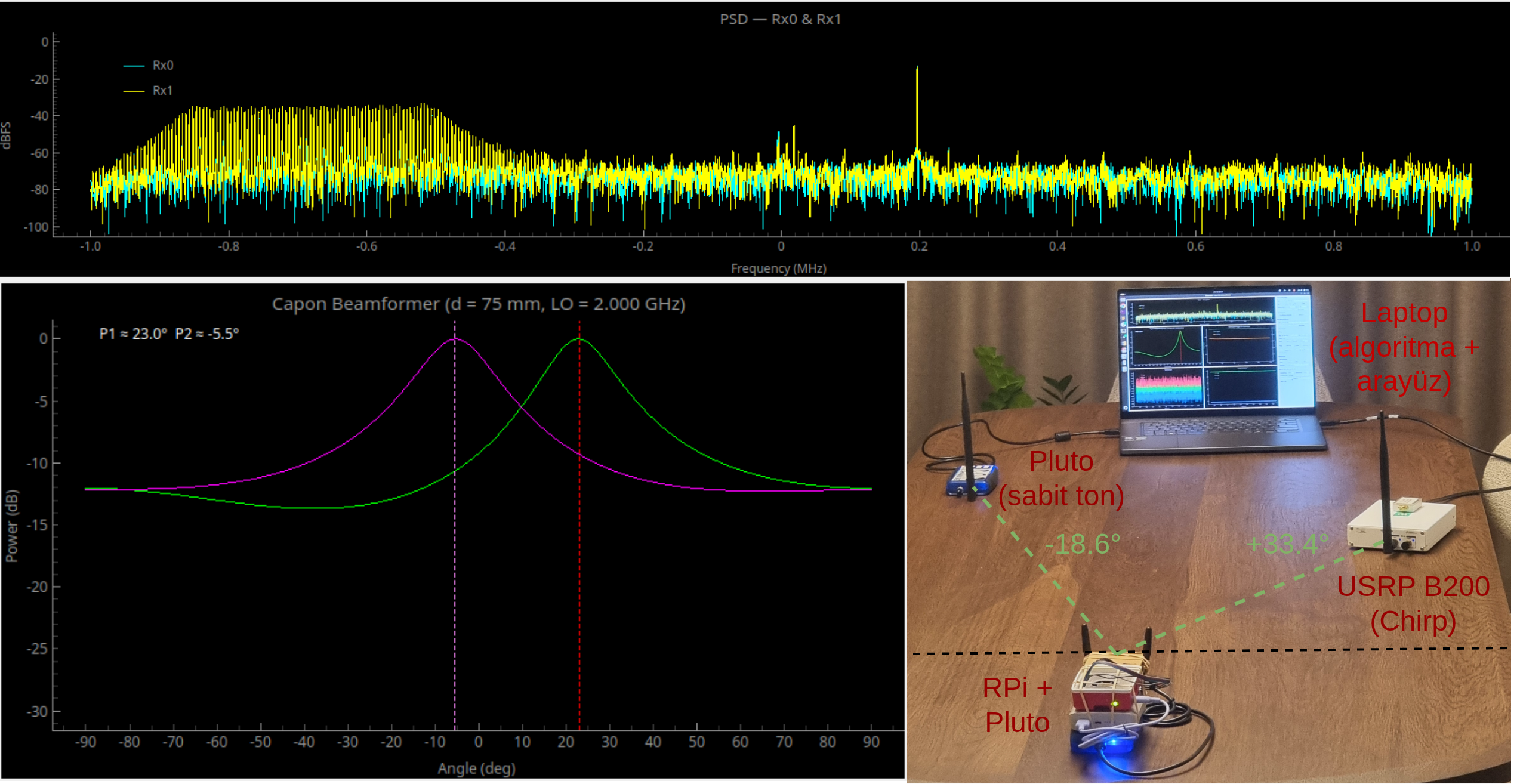}
  \caption{Experimental summary: setup (annotated photograph), two-antenna channel PSD example and Capon spectra. Deviations are observed due to multipath.}
  \label{fig:hwlong_en}
\end{figure*}

For the two-transmitter scenario, the left part of Fig.~\ref{fig:simall_en}~(a) shows the average power (PSD) on the first channel after the filter output. The Tx0 constant tone is around \(\approx -0.4\) normalized frequency, while Tx1, separated from it, is a chirp roughly in the \(0.2\)--\(0.35\) range. The right panel contains the DoA estimates produced by the method. While the true angles are roughly \(-31^\circ\) and \(+23^\circ\), both sources are correct on average. However, Tx1 has a much narrower distribution, whereas Tx0 shows a wider spread due to its geometric position and orientation. In summary, with the proposed method, the root-mean-square error (RMSE) values for the deviation angle from the ULA principal axis are obtained as \(2.61^\circ\) for Tx0 and \(1.56^\circ\) for Tx1. For the oracle method, the same values are \(2.11^\circ\) and \(1.07^\circ\), respectively. The results show that, especially for geometrically more favorably positioned transmitters such as Tx1, the proposed method can attain performance close to that of the oracle method.
  
The three-transmitter scenario, shown in Fig.~\ref{fig:simall_en}~(b), indicates that the method struggles more with three ANF stages than with two, and that DoA estimation errors increase. Fig.~\ref{fig:sim3txanf_en}, under the same conditions, shows the convergence of the three ANFs from initial normalized-frequency offsets of \(\{-0.25,\,0.0,\,0.25\}\) onto the corresponding transmitters on the combined signal. This confirms that the cascaded filter stages lock onto components in the order of their dominance.

In the two-transmitter boundary analysis, the Tx1 chirp is held fixed, while the center frequency of the Tx0 constant tone is brought step by step closer to Tx1. The purpose of this analysis is to measure the effect on DoA estimation performance of the unavoidable mixing of the ANF traces with each other. In the results shown in Table~\ref{tab:boundary_en}, no serious performance drop is observed down to a separation of about 900~kHz, while from 750~kHz onward severe degradations appear, and once the two transmitters overlap in the frequency plane, DoA estimation breaks down entirely. These results indicate the limits the proposed method can reach with the existing simple ANF techniques.

In the ADALM-Pluto-based experimental setup, a variant of the two-transmitter scenario is realized. Indoors, a Pluto with two receive antennas at half-wavelength spacing is connected to a Raspberry Pi unit, which sends the IQ measurements over the local network to the laptop driving the Tx's, where the DoA estimation also runs. The measured PSDs and the corresponding Capon spectra, together with the setup, are shown in Fig.~\ref{fig:hwlong_en}. Although estimates close to and on the same side as the true angles are obtained, deviations in DoA estimates are visible due to the dense multipath in the indoor environment.

\section{Conclusion}

The proposed method enables DoA estimation via the Capon method for \(N\ge 2\) narrowband transmitters on \(M=2\) antennas by separating them through ANFs, and its performance is analyzed. For narrowband signals separable in the time-frequency-angle planes, this method exhibits performance close to that of an oracle method, i.e., a method that knows the signals in advance. For transmitters located at close frequencies and/or under strong multipath, the method begins to suffer degradation due to the structure of ANFs, and a drop in DoA estimation performance is observed. In experimental Pluto measurements, additional degradations due to dense multipath are also visible. In terms of practical applicability, the proposed method is suitable for scenarios in which the frequency separation is on the order of at least a few hundred kHz and the number of transmitters can be estimated. Future work will address these issues with experimentally cleaner field measurements and with algorithmic improvements aimed at mitigating multipath effects.

\clearpage

\selectlanguage{turkish}
\shorthandoff{=}
\renewcommand\tablename{TABLO}
\setcounter{section}{0}
\setcounter{equation}{0}
\setcounter{table}{0}
\setcounter{figure}{0}

\twocolumn[%
  \begin{center}
    {\Large\bfseries Çoklu Sinyal Yön Kestirimi için Adaptif Çentik Filtre Tabanlı Hafif Bir Yöntem\par}
    \vspace{1.2em}
    {\normalsize Burak Soner\par}
    \vspace{0.2em}
    \textit{sobu Labs}\\
    Çankaya, Ankara, Türkiye\\
    burak@sobulabs.com
    \par
  \end{center}
  \vspace{1.2em}
]

\begin{ozet}
Çoklu sinyal tespiti ve yön kestirimi için geleneksel çözüm olan Capon hüzme oluşturma yöntemi, alıcı anten sayısının bir eksiğinden fazla verici içeren senaryolarda bozunum gösterir. Bu bildiride, yalnızca iki alıcı anten ile, iki ve daha fazla dar bant sinyalin aynı anda yön kestirimi için adaptif çentik filtreler (ANF) kullanan hafif hesap yüklü bir yöntem önerilmektedir. Önerilen yöntemde, gelen sinyale ANF katmanları ardışık olarak uygulanarak her bir sinyal için yaratılan izole kanallar üzerinden Capon yöntemi ile yön kestirimi yapılır. Yöntem, oldukça düşük hesap yükü sağlarken, benzetim sonuçlarında ise zaman-frekans-açı düzlemlerinde ayrık vericiler için, sinyalleri önden bilen bir ``kahin'' yönteme benzer başarım göstermektedir. ANF'lerin kendisi gibi bu yöntemin de temel limitasyonu yakın frekanstaki çoklu sinyallerin iyi ayrışamamasıdır. Benzetim sonuçlarının yanı sıra, düşük maliyetli bir yazılım tabanlı radyo platformunda (ADALM-Pluto) deneysel gerçekleme de sunulmuştur. 
\end{ozet}
\begin{IEEEanahtar}
adaptif çentik filtre, yön kestirimi, düşük karmaşıklık, gömülü sistem, faz dizinli sinyal işleme.
\end{IEEEanahtar}

\section{G{\footnotesize İ}r{\footnotesize İ}ş}

Verici konumlama ve takip uygulamalarında faz dizinli antenler vasıtasıyla yön kestirimi ve üçgenleme yöntemleri yaygınca kullanılmaktadır \cite{vantrees2002}. Pratik saha koşullarında sıkça tek bir vericiden ziyade, birbirinden bağımsız ve aynı anda yayın yapan birden çok verici görülebilir. Bu problem uzun süredir faz dizinli sinyal işleme literatüründe ele alınmaktadır. Spesifik olarak, yeterli sayıda anten varlığında Capon hüzme oluşturma \cite{capon1969} ve benzeri yöntemler bu tür çoklu vericilerin yönlerini yüksek çözünürlükle ayırt edebilmektedirler \cite{schmidt1986}.

Buna karşın, bu tür sistemlerin az sayıda anten elemanına sahip düşük maliyetli ve gömülü gerçeklemeleri, özellikle de yakın zamanda yaygınlaşan yazılım tanımlı radyo (SDR) platformları ile birlikte hala güncel bir ihtiyaçtır. Bu bildiride, zaman - frekans - açı (uzay) düzlemlerinde yeterince ayrık dar bant vericiler için (sabit veya frekans-taramalı tonlar), yalnızca iki alıcı anten ile yön kestirimi için adaptif çentik filtre (ANF) tabanlı, hesaplamada hafif bir yöntem önerilmektedir. Bildirinin ana katkıları aşağıdaki şekildedir:

\begin{itemize}
  \item Önerilen yöntem, Fourier dönüşümü tabanlı yöntemlere \cite{wang2014jsee} kıyasla daha düşük hesap yükü sağlamaktadır.
  \item Yöntem, zaman - frekans - açı düzlemlerinde yeterince ayrık vericiler için, sinyalleri önden bilen bir ``kahin'' yönteme benzer başarım göstermektedir.
  \item Yöntem, ADALM-Pluto SDR üzerinde gerçeklenerek deneysel olarak kanıtlanmakta ve analiz edilmektedir.
\end{itemize}

Bu katkıların yanında, ANF'lerin temel zorluğu olan yakın frekanslı çoklu sinyallerin ayrışamamasının bu yöntemin de ana limitasyonu olduğu gösterilmektedir. Gelecek çalışmalar bu problemi çözmeye yönelik olacaktır.

\section{Sistem modeli}

\(M\) alıcı antenli bir düzenek için, \(N\) bağımsız dar bant vericinin birleşimi ve anten başına gürültü eklentisi aşağıdaki gibi modellenir:
\begin{equation}
\mathbf{x}[n] = \sum_{i=1}^{N} \mathbf{a}_i(\theta_i)\, s_i[n] + \mathbf{w}[n],\quad \mathbf{x}[n]\in\mathbb{C}^{M}.
\end{equation}
Burada $\mathbf{x}[n]$ alıcı antenlerde oluşan ana bant (baseband) sinyaller, \(\mathbf{a}_i\) vektörü \(\theta_i\) yönü ile ilişkili yönlendirme vektörü, \(s_i[n]\) ise \(i\) numaralı vericinin ana bant sinyalidir. Gürültü \(\mathbf{w}[n]\) örnekler ve elemanlar arasında beyaz (i.i.d.) varsayılmaktadır.

Dar bant sinyaller ve farklı geliş açılarında yeterince ayrık olan yönlendirme vektörleri için, \(M\) antenli bir düzenekte en fazla \(M{-}1\) vericinin yönü klasik Capon türü yöntemlerle belirlenebilir \cite{capon1969}. Verici sayısı \(N \ge M\) olduğunda ise doğrudan Capon ile ayrımın performansı düşüş gösterir (kestirim, yanlış şekilde iki vericinin yönleri arasında bir ara çözüme yakınsar) \cite{ma2010kr}. Klasik MUSIC ve ESPRIT gibi altuzay yöntemleri de aynı \(N<M\) varsayımına bağlıdır \cite{molaei2024review}. Bu rejimde, bu makalede önerilen yöntemdeki gibi bir frekans-zaman düzlemi ayrıştırması ile yön kestirimi performansını hafif hesap yüküyle iyileştirmek mümkündür.

\section{Önerilen Yöntem}

Bu bölümde önce verici ayrıştırma için kullanılan IIR ANF yapısı tanıtılır, ardından ANF ile kurulan ardışık verici izolasyon aşamaları anlatılır, ve son olarak da her izole kanal üzerinde uygulanan, Capon yöntemi vasıtasıyla gerçeklenen verici yön kestirimi yaklaşımı açıklanır. Yöntemin önemli bir öncülü, alınan sinyalde kaç adet verici olduğunu önden bilindiğini varsaymasıdır. Bu varsayım, mevcut ardışık ANF yapısına yapılabilecek basit bant gücü tanıma eklemeleriyle ortadan kaldırılabilir, fakat bu özellik yer kısıtlarından ötürü makalenin konusunun dışında tutulmuştur.

\subsection{Adaptif Çentik Filtresi (ANF)}

Karmaşık (complex) dar bant sinyallerin takibi için ANF yapıları literatürde yaygınca bilinmektedir \cite{nehorai1985}. Pratikte düşük hesap yükü nedeniyle tek kutuplu IIR çentik filtre yapısı ve gradyan tabanlı optimizasyon yöntemleri tercih edilmiştir \cite{borio2006ion}. $S$ ANF aşaması ve $M$ anten için örnek başına $\mathcal{O}(SM)$ ile çalışan ANF'ler, örnek başına en az $\mathcal{O}(M\log N)$ ile çalışan FFT-tabanlı yöntemlere göre kısıtlı kaynaklı donanımlarda gerçek zamanlı çalışmaya daha elverişlidir \cite{trogliagamba2012}. Bu tür bir ANF, yeterince ayrık dar bantlı çoklu bileşenlerin (sabit ton ve chirp) anlık frekanslarını örnekten örneğe takip eder, ve her ardışık aşamada baskın bileşene odaklanır. 

Önerilen yöntemde tek bir anten için kullanılan ANF yapısı, her anten ve her ardışık ANF aşaması için aynen tekrarlanır. Karmaşık örnek \(x_n\), filtre çentik frekansı \(\omega_n\) ve kutup yakınlaştırma faktörü \(k_a\) ile filtrenin ara durumu \(x^r_n\) ve filtre çıktısı \(y_n\) şu biçimdedir:
\begin{equation}
z_n = e^{j\omega_n},\quad
x^r_n = x_n + k_a z_n x^r_{n-1},\quad
y_n = x^r_n - z_n x^r_{n-1}.
\end{equation}
Normalize en küçük ortalama kareler algoritması (NLMS) ile güncellenen \(\omega_n\), şu denklemleri kullanır \cite{haykin2013}:
\begin{equation}
\omega_{n+1} = \mathrm{clip}\!\left(
\omega_n - \frac{\mu}{\hat{P}_n}\,
\mathrm{Re}\!\left\{ y_n^* \, j z_n x^r_{n-1} \right\}
\right),
\end{equation}
burada \(\hat{P}_n\) üstel hareketli ortalama (EMA) ile kestirilmiş giriş gücüdür, NLMS kazancı \(\mu\) pozitif bir sabittir, ve clip operasyonu ise Nyquist sınırlarına göre yapılan sınırlamadır.

\subsection{Sinyal Birleştirme ve Verici İzolasyonu}

Önerilen yöntemde her anten \(k\) için ana bant (baseband) \(x_k[n]\) üzerinde \(S\) aşamalı ardışık ANF çalıştırılır. Bir aşamanın çıkışı, aynı antenin bir sonraki aşamasının girdisidir. Her aşamada elemanlar arasından elde edilen anlık frekans izleri \(\hat{f}_k[n]\) (Hz), örnek bazında aritmetik ortalama ile birleştirilir:
\begin{equation}
\bar{f}[n] = \frac{1}{M}\sum_{k=1}^{M} \hat{f}_k[n].
\end{equation}
Bileşik izler \(\bar{f}^{(s)}[n]\), \(s=1,\ldots,S\), uyarlama bittikten sonra aynı sabit çentik filtre yapısı ile her antene uygulanır. Fiziksel olarak hangi vericinin hangi aşamaya denk geldiği rastgeledir, ve sıralama örtüşmeyebilir. Hangi bileşenin önce yakalandığı baskınlığa ve mevcut ANF'nin o anda o banda yakınlığına bağlıdır. Her hedef verici \(j\in\{1,\ldots,S\}\) için, gözlenen ham çok kanallı anlık sinyal \(\mathbf{X}\) üzerinde, \(k\neq j\) olan tüm aşamaların birleşik izleriyle çentik filtreler uygulanır, ve böylece \(j\) vericisine ait izole sinyal kanalı elde edilir.

\subsection{Yön Kestirimi}

İzole kanallar elde edildikten sonra, her kanal için $M\times M$ boyutunda örnek kovaryans matrisi \(\hat{\mathbf{R}}_j\) hesaplanır. İki elemanlı ULA için yönlendirme vektörü \(\mathbf{a}(\theta)\) altında Capon formülasyonu aşağıdaki gibidir \cite{capon1969}:
\begin{equation}
P(\theta) = \frac{1}{\mathbf{a}(\theta)^H (\hat{\mathbf{R}}_j)^{-1} \mathbf{a}(\theta)}.
\end{equation}
Bu yöntem ile üretilen Capon sinyalindeki en yüksek değere (tepeye) denk gelen açı değeri, yön kestirimi olarak kullanılır.

\section{Benzetim ve Deneysel Sonuçlar}

Bildiride benzetim, hem açık alan hem de iç mekanlarda gerçekçiliği korumak adına yol gecikmesi ve çoklu yol (multipath) etkisini hesaba katar. Bunun için dikdörtgen bir hacim içinde tek sekme (bounce) etkisini modeller. Tablo~\ref{tab:simenv}'de özetlenen ortamda iki elemanlı ULA merkezi \((8.2, 2.8, 1.2)\,\mathrm{m}\) konumunda, eleman aralığı \(\lambda/2\) (\(f_c=2\,\mathrm{GHz}\)), ana bant örnekleme \(f_s=4\,\mathrm{MHz}\), anlık örnek uzunluğu \(N=512\), anten başına gürültü \(\sigma=0.12~\text{pu}\) (benzetimde kullanılan ölçekte), ANF parametreleri \(k_a=0.70\), \(\mu_{\mathrm{norm}}=0.1\), EMA güç kestirimi için \(\alpha=0.01\) ve altı yüzeyli duvarlar üzerinde rastgele fazlı yansıma katsayıları \(0.33\)--\(0.52\) olarak kullanılmıştır. Tüm benzetim sonuçları \(400\) adet rastgele Monte Carlo denemesi üzerinden elde edilmiştir. İki temel benzetim senaryosu ve ek olarak yakın frekanslı bileşenler altında bozunumu ölçen bir sınır analizi incelenmiştir: 

\begin{table}[b]
  \vspace{10px}
  \centering
  \caption{Benzetim ortamının başlıca parametreleri}
  \label{tab:simenv}
  \begin{tabular}{|l|l|}
    \hline
    Büyüklük & Değer \\
    \hline
    Oda boyutları (m) & \(20\times 12\times 3\) \\
    Taşıyıcı frekans \(f_c\) & \(2\,\mathrm{GHz}\) \\
    Örnekleme \(f_s\), örnek sayısı & \(4\,\mathrm{MHz}\),  \(512\) \\
    ULA eleman aralığı & \(\lambda/2\) \\
    Gürültü (anten başına) & \(\sigma=0.12\) \\
    ANF (\(k_a\), \(\mu_{\mathrm{norm}}\), \(\alpha\)) & \(0.70\), \(0.1\), \(0.01\) \\
    Çok yol & Duvardan tek sekme, rastgele faz \\
    \hline
  \end{tabular}
\end{table}

\begin{enumerate}
  \item \emph{İki verici:} zaman-frekans-açı düzlemlerinde yeterince ayrık iki kaynak için sonuçlar, ve kahin yöntem ile karşılaştırma.
  \vspace{3px}
  \item \emph{Üç verici:} aynı deneyin üç verici ile tekrarı, ve ayrıklığın azalmasının negatif etkisinin gözlemi. 
  \vspace{2px}
  \item \emph{Sınır analizi:} önerilen yöntemin limitlerini belirleyen, iki vericinin birbirine yaklaşmasıyla ANF izlerinin birbirine karışması, ve yön kestirimlerinin kötüleşmesinin analizi.
\end{enumerate}

\begin{figure*}[t]
  \centering
  \includegraphics[width=0.94\textwidth]{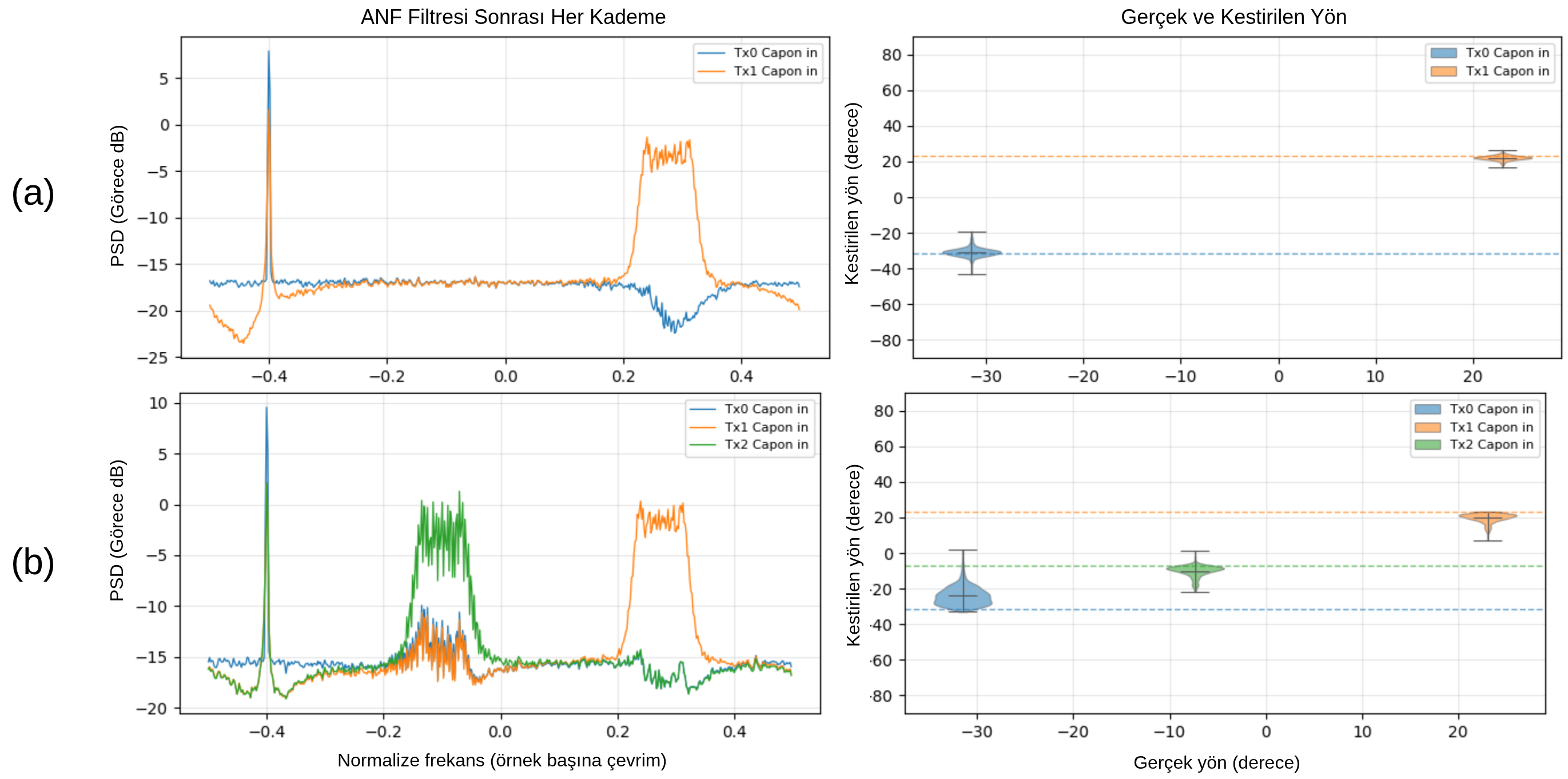}
  \caption{Benzetim özeti: (a) İki verici: filtre çıkışı sonrası kanal PSD'si ve Monte Carlo yön dağılımları (Tx1 yön kestirimi Tx0'a göre daha isabetli), (b) Üç ayrık verici: ANF+Capon çıktıları.}
  \label{fig:simall}
\end{figure*}

\begin{figure}[t]
  \centering
  \includegraphics[width=0.92\columnwidth]{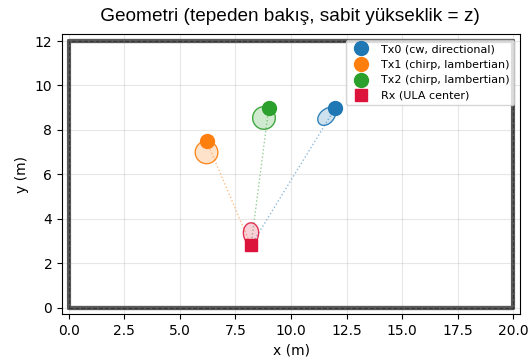}
  \caption{Benzetim geometrisi (üstten görünüm): üç verici, ULA merkezi ve oda sınırları. İki vericili senaryolarda Tx2 devre dışıdır.}
  \label{fig:simsetup}
  \vspace{4px}
\end{figure}

\begin{figure}[t]
  \centering
  \includegraphics[width=0.92\columnwidth]{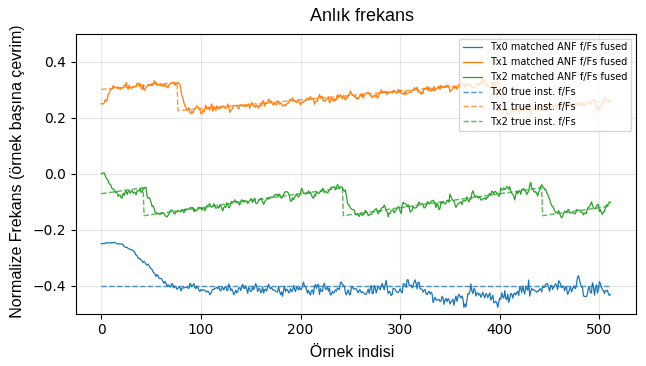}
  \caption{Üç kademeli ANF yapısının anlık frekans izi. Başlangıç ofsetlerinden sonra izler fiziksel vericilere karşılık gelen frekans bölgelerine oturuyor.}
  \label{fig:sim3txanf}
  \vspace{7px}
\end{figure}

\begin{table}[t]
  \centering
  \caption{Frekansları yaklaşan iki vericide sınır analizi}
  \label{tab:boundary}
  \begin{tabular}{|r|r|r|}
    \hline
    $\Delta f$ (kHz) & Tx0 RMSE & Tx1 RMSE \\
    \hline
    $-2500$ & 4.44 & 1.50 \\
    $-1200$ & 4.70 & 1.68 \\
    $-900$ & 5.56 & 2.26 \\
    $-750$ & 15.12 & 3.30 \\
    $-600$ & 24.66 & 6.44 \\
    $-200$ & 26.79 & 5.58 \\
    \multicolumn{1}{|l|}{Örtüşme} & 35.23 & 6.66 \\
    \hline
  \end{tabular}
\end{table}

Şekil~\ref{fig:simsetup} benzetim düzleminde üç verici (Tx0--Tx2), iki elemanlı ULA alıcısı ve tek sekme çok yol geometrisini üstten göstermektedir. Tüm vericiler aynı \(z\) yüksekliğinde tutularak iki antenli ULA ile yalnızca yanca (azimuth) yön kestirimi hedeflenmiştir. Tüm vericilerin anten hüzmeleri şekilde eliptik (yönlü) ve yuvarlak (Lambertian) olarak gösterilmiştir. İki vericili deneylerde Tx2 kapatılarak yalnızca Tx0 ve Tx1 kullanılmıştır.

\begin{figure*}[t]
  \centering
  \includegraphics[width=0.90\textwidth]{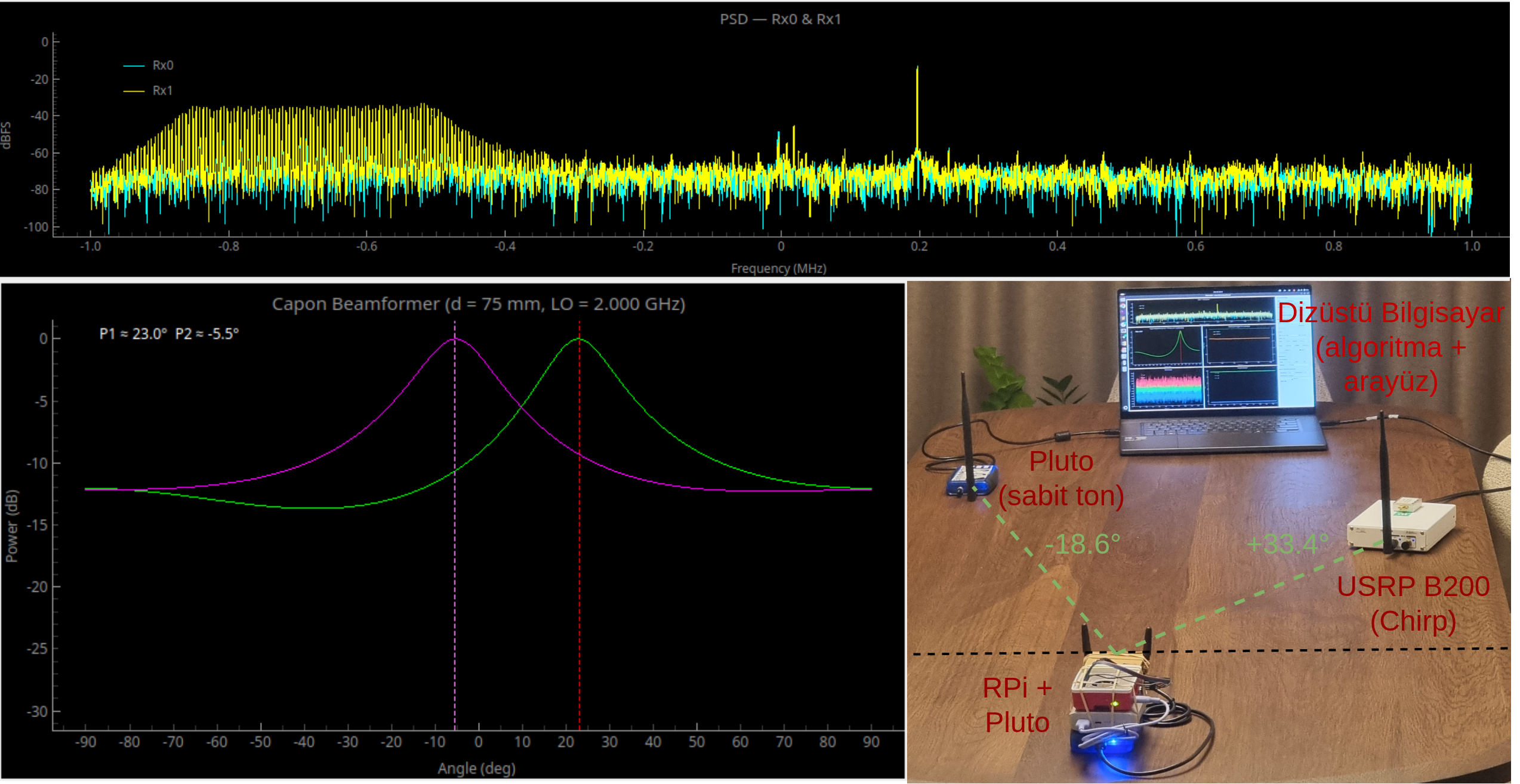}
  \caption{Deneysel özet: kurulum (işaretli fotoğraf), iki anten kanalı PSD örneği ve Capon spektrumları; çok yol nedeniyle sapmalar gözlenmektedir.}
  \label{fig:hwlong}
\end{figure*}

İki vericili senaryo için Şekil~\ref{fig:simall}~(a) sol bölümde, filtre çıkışı sonrası ilk kanal üzerinde ortalama güç (PSD) gösterilmektedir. Tx0 sabit tonu \(\approx -0.4\) normalize frekans civarındadır. Tx1 ise ondan ayrık, yaklaşık \(0.2\)--\(0.35\) aralığında chirp formundadır. Sağ panelde yöntemin yön kestirimleri yer almaktadır. Gerçek açılar kabaca \(-31^\circ\) ve \(+23^\circ\) iken, her iki kaynak da ortalamada doğrudur, ancak Tx1 çok daha dar, Tx0 ise geometrik konum ve yönelimden ötürü daha geniş bir yayılım gösterir. Özetle önerilen yöntem ile ULA ana eksenden sapma açısı kök ortalama kare hata (RMSE) değerleri Tx0 için \(2.61^\circ\), Tx1 için \(1.56^\circ\) olarak elde edilmiştir. Kahin yöntem için aynı değerler sırasıyla \(2.11^\circ\) ve \(1.07^\circ\) dir. Sonuçlar, özellikle geometrik olarak daha avantajlı konumlanmış Tx1 gibi vericiler için, önerilen yöntemin kahin yönteme yakın başarım gösterebildiğini ortaya koymaktadır.
  
Üç vericili senaryo Şekil~\ref{fig:simall}~(b) üzerinde, yöntemin üç ANF aşamasında ikiye göre daha çok zorlandığını, ve yön kestirimi hatalarının arttığını göstermektedir. Şekil~\ref{fig:sim3txanf} ise aynı koşulda üç ANF'nin \(\{-0.25,\,0.0,\,0.25\}\) normalize frekans başlangıçlarından, birleştirilmiş sinyal üzerinden ilgili vericilere yakınsamasını göstermektedir. Bu, ardışık filtre aşamalarının baskın bileşen sırasına göre kilitlendiğini doğrular.

İki vericili sınır analizinde Tx1 chirp'i sabit tutulmuş, Tx0 sabit tonunun merkez frekansı adım adım Tx1'e yaklaştırılmıştır. Bu analizdeki amaç ANF izlerinin birbirine kaçınılmaz şekilde karışmasının yön kestirim performansına etkisini ölçmektir. Tablo~\ref{tab:boundary} ile gösterilen sonuçlarda yaklaşık 900 kHz'lik bir farka kadar performansta ciddi bir düşüş görülmezken, 750 kHz'ten itibaren ciddi bozunumlar görülmekte, iki verici frekans düzleminde örtüştüğü zaman ise yön kestirimi tamamen bozulmaktadır. Bu sonuçlar önerilen yöntemin mevcut basit ANF teknikleri ile erişebildiği sınırları göstermektedir.

ADALM-Pluto tabanlı deneysel düzenekte iki vericili senaryonun bir çeşidi canlandırılmıştır. İç mekanda bir Raspberry Pi ünitesine bağlı, yarım dalgaboyu aralıklı iki alıcı antenli Pluto, yerel ağ üstünden Tx'leri süren laptop'a IQ ölçümlerini göndermekte, ve yön kestirimi laptop'ta koşmaktadır. Ölçülen PSD'ler ve ilgili Capon spektrumları, düzenek ile beraber Şekil~\ref{fig:hwlong} ile gösterilmiştir. Esas açılara yakın ve aynı yönde kestirimler elde edilmiş olsa da, iç mekandaki yoğun çok yol (multipath) nedeniyle yön kestirimlerinde sapmalar gözlenmektedir.

\section{Sonuç}

Önerilen yöntem, \(M=2\) antende \(N\ge 2\) dar bant vericinin Capon yöntemi ile yön kestirimi yapılabilmesi için, ANF vasıtasıyla ayrıştırılmasını önerir ve performansını analiz eder. Zaman-frekans-açı düzlemlerinde ayrık olan dar bant sinyaller için bu yöntem, kahin bir yönteme, yani sinyalleri önden bilen bir yönteme yakın bir performans sergiler. Yakın frekanslarda bulunan vericilerde ve/veya yüksek çok yol (multipath) etkilerinde ise yöntem ANF'lerin yapısından ötürü bozunum yaşamaya başlar, ve yön kestirim performansında düşüş görülür. Deneysel Pluto ölçümlerinde ek olarak yoğun multipath kaynaklı bozunumlar görülebilmektedir. Pratik uygulanabilirlik açısından önerilen yöntem, frekans ayrıklığı en az birkaç yüz kHz mertebesinde olan ve verici sayısının tahmin edilebildiği senaryolar için uygundur. Gelecekteki çalışmalarda deneysel olarak daha temiz saha ölçümleri ve çok yol etkilerinin giderimine yönelik algoritma iyileştirmeleri ile bu sorunlar ele alınacaktır.


\begin{thebibliography}{99}

\bibitem{vantrees2002}
H. L. Van Trees, \emph{Detection, Estimation, and Modulation Theory, Part IV: Optimum Array Processing}. New York, NY, USA: Wiley, 2002.  
  
\bibitem{capon1969}
J. Capon, ``High-resolution frequency-wavenumber spectrum analysis,'' \emph{Proc. IEEE}, vol.~57, no.~8, pp.~1408--1418, Aug. 1969.

\bibitem{schmidt1986}
R. O. Schmidt, ``Multiple emitter location and signal parameter estimation,'' \emph{IEEE Trans. Antennas Propag.}, vol.~34, no.~3, pp.~276--280, Mar. 1986.

\bibitem{wang2014jsee}
X. Wang, Z. Huang, and Y. Zhou, ``Underdetermined DOA estimation and blind separation of non-disjoint sources in time-frequency domain based on sparse representation method,'' \emph{J. Syst. Eng. Electron.}, vol.~25, no.~1, pp.~17--25, Feb. 2014, doi:~10.1109/JSEE.2014.00003.

\bibitem{ma2010kr}
W.-K. Ma, T.-H. Hsieh, and C.-Y. Chi, ``DOA estimation of quasi-stationary signals with less sensors than sources and unknown spatial noise covariance: A Khatri--Rao subspace approach,'' \emph{IEEE Trans. Signal Process.}, vol.~58, no.~4, pp.~2168--2180, Apr. 2010, doi:~10.1109/TSP.2009.2034935.

\bibitem{nehorai1985}
A. Nehorai, ``A minimal parameter adaptive notch filter with constrained poles and zeros,'' \emph{IEEE Trans. Acoust., Speech, Signal Process.}, vol.~ASSP-33, no.~4, pp.~983--996, Aug. 1985.

\bibitem{borio2006ion}
D. Borio, ``Analysis of the one-pole notch filter for interference mitigation: Wiener solution and loss estimations,'' in \emph{Proc. ION GNSS}, Fort Worth, TX, USA, Sep. 2006.

\bibitem{trogliagamba2012}
M.~Troglia Gamba, E. Falletti, D. Rovelli, and A. Tuozzi, ``FPGA implementation issues of a two-pole adaptive notch filter for GPS/Galileo receivers,'' in \emph{Proc. ION GNSS}, Nashville, TN, USA, Sep. 2012, pp.~3549--3557.

\bibitem{haykin2013}
S. Haykin, \emph{Adaptive Filter Theory}, 5th ed. Upper Saddle River, NJ, USA: Pearson, 2013.

\bibitem{molaei2024review}
A. M. Molaei, B. Zakeri, S. M. Hosseini Andargoli, M. A. B. Abbasi, V. Fusco, and O. Yurduseven, ``A comprehensive review of direction-of-arrival estimation and localization approaches in mixed-field sources scenario,'' \emph{IEEE Access}, vol.~12, pp.~65883--65918, 2024, doi:~10.1109/ACCESS.2024.3398351.

\end{thebibliography}
\end{document}